\documentclass[english,aps,prl,onecolumn,amsmath,amssymb, 
showkeys,notitlepage]{revtex4-1}

\usepackage{setspace}
\usepackage{xkeyval}

\usepackage[normalem]{ulem}
\usepackage{graphicx}

\usepackage{bm}

\usepackage[hidelinks=true]{hyperref}

\hypersetup{colorlinks=true,
	    citecolor=blue,
	    linkcolor=blue,
            urlcolor = blue,	   
}

\usepackage{xcolor}
\usepackage{titlesec}

\usepackage{soul}

\newcommand{\rv}{$\delta R^2 / R^2$ }
\newcommand{\res}{$R$ }

\begin{document}

\title{High performance sensors based on resistance fluctuations of single layer graphene transistors}

\author{Kazi Rafsanjani Amin}
\author{Aveek Bid}
\email{aveek.bid@physics.iisc.ernet.in}
\affiliation{Department of Physics,Indian Institute of Science, Bangalore,\\ Karnataka, India 560012}

\keywords{graphene, field effect transistor, sensor, resistance fluctuations, noise, number-density fluctuation}

\begin{abstract}
One of the most interesting predicted applications of graphene monolayer based devices is as high quality sensors. In this letter we show, through systematic experiments, a chemical vapor sensor based on the measurement of low frequency resistance fluctuations of single layer graphene field-effect-transistor (SLG-FET) devices. The sensor has extremely high sensitivity, very high specificity, high fidelity and fast response times. The performance of the device using this scheme of measurement (which uses resistance fluctuations as the detection parameter) is more than two orders of magnitude better than a detection scheme where changes in the average value of the resistance is monitored. We propose a number-density fluctuation based model to explain the superior characteristics of noise measurement based detection scheme presented in this article.

\end{abstract}

\maketitle

\section{INTRODUCTION}

Single layer graphene (SLG) has several distinctly unique properties that make it exceptionally suited for use as material and radiation sensors. The specific surface area (2630 $m^2/g$) of SLG is amongst the highest in layered materials~\cite{Pumera2010954} making the conductance of graphene extremely sensitive to the ambient - the presence of a few foreign molecules on its surface can significantly modify its electrical characteristics. SLG is highly conductive even in very low carrier density regimes with room temperature mobilities of the order of 20,000 $cm^2/Vs$ routinely achievable~\cite{Soldano20102127,zheng:063110,2010AdPhy..59..261A}. This  causes graphene based devices to have much lower levels of  thermal noise as compared to semiconductor based sensors having comparable carrier densities. The low defect levels of pristine graphene ~\cite{Novoselov22102004,doi:10.1021/nl080241l,doi:10.1021/es902659d,4773243} ensures that intrinsic flicker noise due to thermal switching of defects are lower than any semiconductor material \cite{balandin2013low, xu2010effect, pal2011microscopic, kaverzin2012impurities, pellegrini20131}. The ability of SLG to interact with materials with a variety of interactions - from weak van der Waals force to extremely stable covalent bonds raises the possibility of detecting a wide variety of materials with SLG based sensors with high specificity. Single layer graphene field-effect-transistor (SLG-FET) devices thus seem to have almost all the properties required to be an effective sensor material - accessibility to large surface area, good transduction, electrical and mechanical stability and ease of preparation.

There have been previous reports of the use of graphene based sensors  to detect various chemical gas molecules like $NH_3$, CO, $NO_2$, NO, $O_2$, $CO_2$ and $H_2$~\cite{schedin2007detection, chen:053119, chen:243502,Yoon2011310, Wu2010296, amin2014graphene, yavari2012graphene, yuan2013graphene,:/content/aip/journal/apl/103/5/10.1063/1.4816762},
as well as  bio-molecules~\cite{ADMA:ADMA200903645,dong2011situ}.
In all these cases the change in resistance of the device was used as the detection parameter. The sensitivity obtained was at best a few percentage with extremely long device reset times (of the order of tens of minutes to hours) making them unsuitable for any practical applications. This scenario motivates the development of alternate schemes of sensing, which allows fast detection of analytes with similar, if not improved, sensitivity. In a previous publication we had reported a very high sensitivity of the intrinsic low frequency resistance noise of SLG-FET devices to the nature of its ambient. We had also elucidated  a probable mechanism behind the high sensitivity of the measured noise to changes in the ambient of the graphene device ~\cite{:/content/aip/journal/apl/106/18/10.1063/1.4919793}.
In this letter we present extensive studies of the sensing of specific gas molecules using resistance fluctuations (noise) of SLG-FET devices. The relative variance \rv of resistance fluctuations of the devices were found to show reproducible changes upon exposure to many different chemical vapor molecules. The devices had extremely fast response and reset times, of the order of  seconds. The sensitivity of the SLG FET sensor using this technique was more than an order of magnitude better than sensing with the same device using changes in the resistance of the device. 

There has been previous demonstrations of the use of resistance fluctuations to detect adsorbed molecules \cite{doi:10.1021/nl3001293} but a systematic study of the sensitivity, specificity and response times of the sensors based on this technique is missing. To work as an effective material sensor a device must satisfy a certain basic set of criterion: (1) it should have a measurable response, (2) the response times and reset times must be low, (3) its response must scale as the amount of test molecules in its working range, (4) the response must be reproducible  and (5) there should be selectivity in response to different types of test molecules. We show in this letter that SLG-FET using resistance fluctuations as the detection parameter satisfies all the above criterion extremely well.


\section{MEASUREMENT}

\begin{center}
\begin{figure}[!th]
\includegraphics[width = 0.5\textwidth]{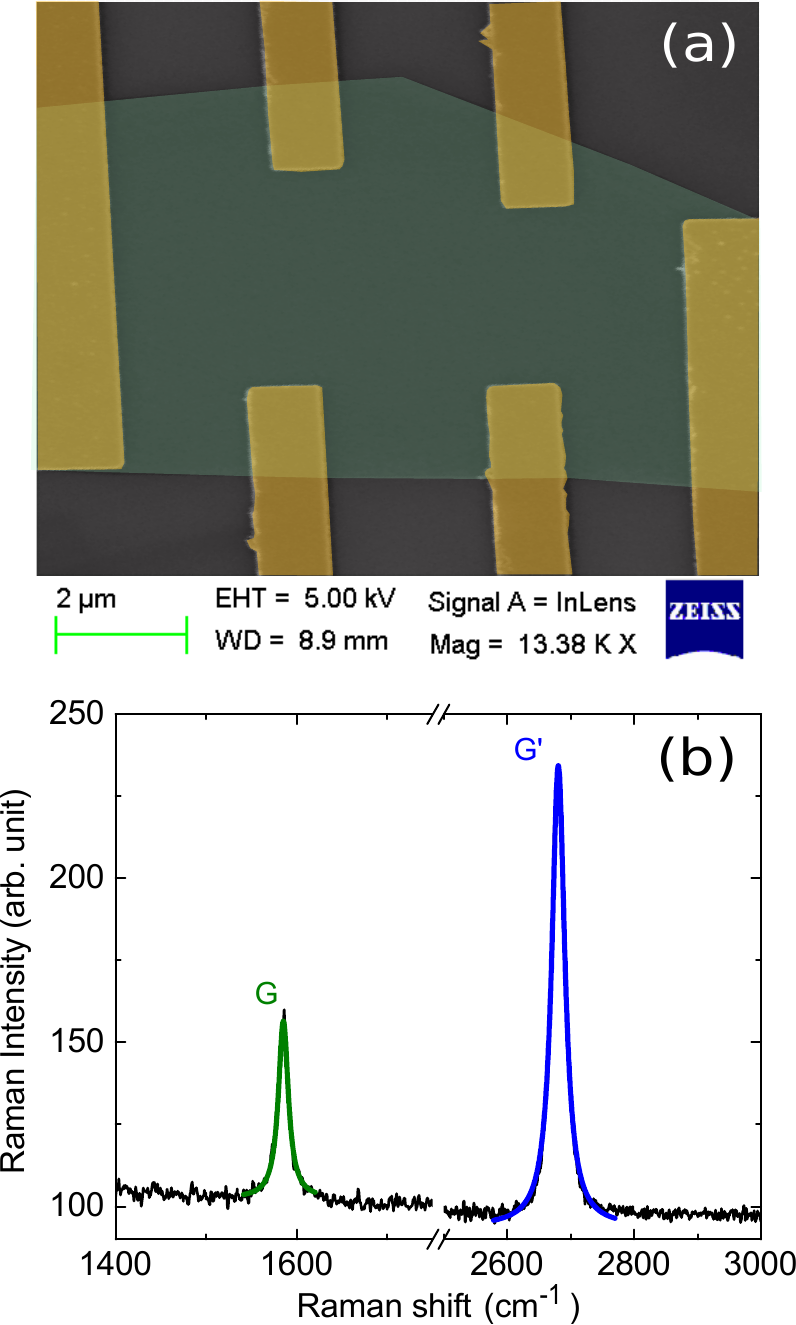}
    \small { \caption{ (a) False color SEM image of a typical graphene device. The graphene sheet is shown in light green and the metal contact pads are shown in yellow. (b) Raman spectra of the SLG measured after  the lithography process. The data is shown in black line, where as the green and blue lines are the lorentzian  fits to the G peak and the G$^\prime$ peaks respectively. \label{fig:semraman}}}
\end {figure}
\end{center}

\begin{center}
\begin{figure}[!th]
\includegraphics[width = 0.5\textwidth]{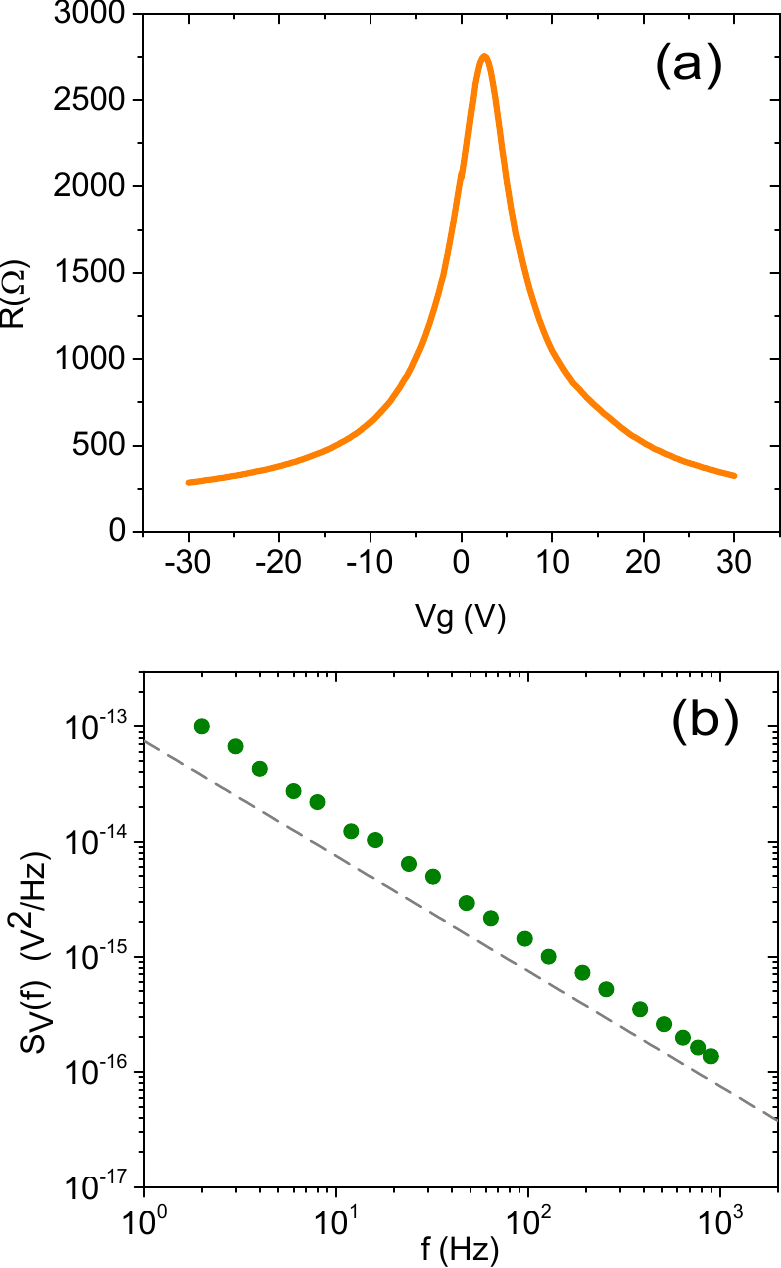}
    \small { \caption{(a) Gate voltage ($V_g$) dependence of resistance (R) of an  SLG-FET device.  (b) A typical 1/f noise power spectrum (olive filled circles)  of a pristine graphene monolayer FET device measured at room temperature. The grey  solid line shows  a reference 1/f curve. \label{fig:rvgnoise}}}
\end {figure}
\end{center}

The devices reported in this letter were prepared from natural graphite exfoliated on $Si/SiO_2$ wafers. Electrical contacts were fabricated using conventional electron beam lithography\cite{Novoselov22102004} followed by thermal deposition of Cr/Au ($5 \sim 7nm / 70 nm$). Atomic force microscopy (AFM) and Scanning electron microscope (SEM) imaging were used to determine the surface quality of devices after the lithography processes. Figure~\ref{fig:semraman}(a) shows a false color  SEM image of a typical SLG-FET device. The graphene sheet is shown in light green and the electrical contact pads are shown in yellow. The number of layers in each device was confirmed through measurements of the Raman spectra of devices~\cite{PhysRevLett.97.187401, ferrari2007raman, raman} and in some cases based on the position of conductance plateaus in the integer quantum hall regime. Figure~\ref{fig:semraman}(b) shows a Raman spectra of the device - the blue and green lines are the Lorentzian fits to the experimental data. The absence of a D-peak, the position of the G peak (1582.2 $cm^{-1}$) and the ratio of the heights of the G-peak to the G$^\prime$-peak in the Raman spectra indicate that there is negligible extrinsic doping in the device. Figure~\ref{fig:rvgnoise}(a) shows the gate voltage $V_g$ dependence of the resistance from which the mobility $\mu$  and the impurity charge carrier concentration $n_0$ of the device were extracted~\cite{PhysRevB.80.235402}. Room temperature mobilities of our typical device were in the range of 10,000-20,000 $cm^2 V^{-1}s^{-1}$ while $n_0$ was about $10^{12}$~cm$^{-2}$. The location of the Dirac point ($V_D$) very close to zero $V_g$ and the low value of $n_0$ both attest to the high quality of the devices. 

\begin{center}
\begin{figure}[!th]
\includegraphics[width = 0.5\textwidth]{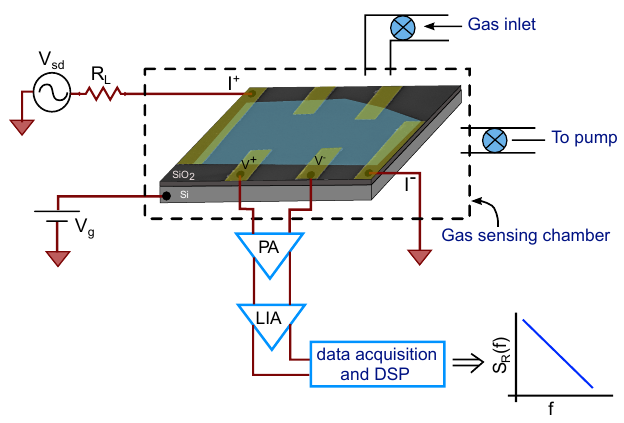}
    \small { \caption{ A schematic diagram of the measurement setup. The gas sensing chamber containing the device is shown by the black dotted line. $V_{sd}$ is the ac source-drain bias voltage, $R_L$ is the series ballast resistance (typically about 1~M$\Omega$) and $V_g$ is the dc gate bias.  PA represents the low-noise pre-amplifier (SR 552) and LIA is the dual channel lock-in amplifier (SR 830). \label{fig:setup}}}
\end {figure}
\end{center}

For electrical measurements the devices were wire-bonded to a lead-less chip carrier, and were mounted to a mating socket, which was wired into a custom-build vacuum chamber. The chamber was made of thick stainless steel for better electrical shielding. A schematic diagram of the measurement set-up is shown in figure~\ref{fig:setup}.
The resistance (\res) of the graphene FET devices was measured by standard low frequency ac lock-in measurement techniques.  
The power spectral density (PSD) $S_R(f)$ of low frequency resistance fluctuations  were measured over the frequency bandwidth of 1~Hz to 1~KHz using an ac lock-in detection technique~\cite{ghosh2004set,scofield1987ac}. For the measurement of noise, the device was biased with a small ac voltage at a carrier frequency significantly higher than the upper cut-off frequency of the noise measurement bandwidth. The mean 4-probe voltage across the device was digitally offset and the voltage fluctuations $\delta V(t)$ about this mean value were recorded from the output channels of a lock-in amplifier using a high speed 16-bit Digital-to-Analog conversion (DAQ) card. The data acquisition rate was determined by the Nyquist criterion which states that the minimum value of the sampling rate must be at least twice the highest frequency spectral component present in the signal being studied
\cite{shannon1949communication}.
The sampling rate was kept usually 8 or 16 times higher than the limit given by Nyquist's sampling theorem, and the time series of resistance fluctuations were recorded in contiguous segments of 30 seconds each.  The  acquired  data was  anti-aliased digitally and down-sampled. The power spectral density of resistance fluctuations (PSD) $S_R(f)$ was estimated using Welch's averaged periodogram method\cite{welch1967use}.
 This technique of noise measurement~\cite{ghosh2004set,scofield1987ac} allows for a simultaneous measurement of both the intrinsic resistance fluctuations of the device and the background noise arising from thermal fluctuations as well as instrumentation noise. A typical PSD, $S_R(f)$ of resistance fluctuations, of the SLG-FET measured at room temperatures is shown in figure~\ref{fig:rvgnoise}(b) as a function of the frequency $f$. The resistance fluctuation spectra of pristine graphene devices were always found to be $1/f$ in nature over the entire bandwidth of measurement.  

The PSD $S_R(f)$ integrated over the frequency bandwidth of measurement and normalized by the square of the average resistance value gives the relative variance of the resistance fluctuations \rv (which we refer to as noise):
\begin{equation}
\frac{\delta R^2}{ R^2} =\frac{1}{R^2} \int^{f_{max}}_{f_{min}}f  S_R(f) df
\label{eqn:rv}
\end{equation}
Here $f_{min}$ and $f_{max}$ are respectively the lower and upper bounds of the measurement bandwidth.

Graphene has a small but finite work-function difference with metal contact probes and the fluctuations in the Fermi level near the contact region can generate measurable resistance fluctuations. To address this issue, we have measured $\delta R^2 / R^2$ of device at different applied source-drain $V_{sd}$. We find that the measured noise always scales with the square of the applied $V_{sd}$ (to within $\pm$5\%) showing that the contribution of the contact noise to the observed effect is negligible 
\cite{:/content/aip/journal/apl/106/18/10.1063/1.4919793}.
 As discussed later in this letter, we also find that the measured noise has a strong dependence on the gate voltage $V_g$ suggesting that the major contribution to the measured noise arises from the bulk of the device and not at the contacts.

\section{RESULTS AND DISCUSSION}

\begin{center}
\begin {figure*}[!tbh]
   \includegraphics[width=0.95\textwidth]{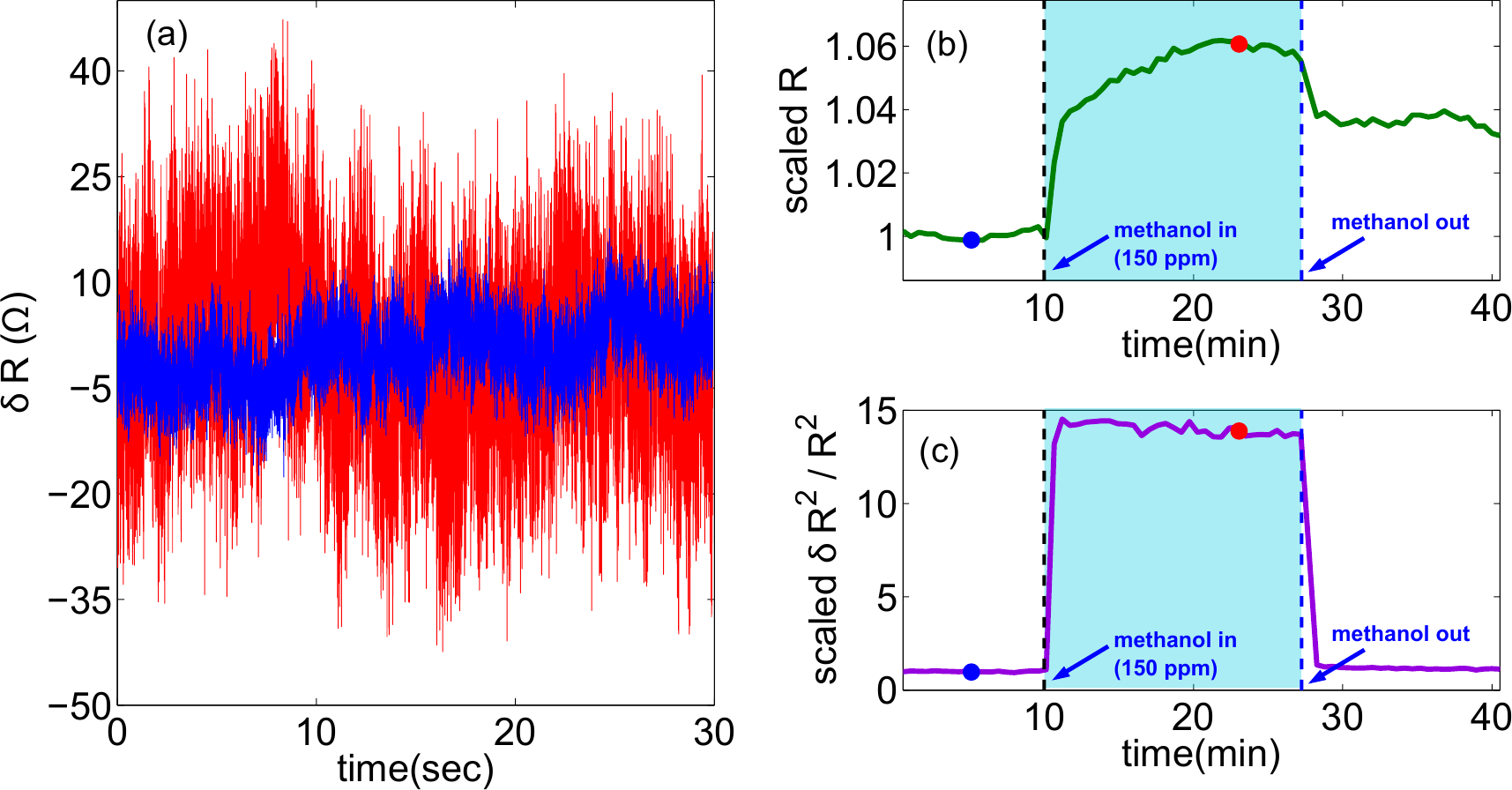}
    \small { \caption{Comparison of the resistance and resistance noise of the SLG-FET device in the absence and presence of analytes. (a) Plot of the time-series of resistance fluctuations about the mean value in the pristine SLG-FET device (blue line) and for the same device in the presence of 150~ppm of methanol (red line). It can be seen that the fluctuations in the device resistance are greatly enhanced in the presence of chemical vapour. (b) Plot of scaled $R$ as a function of time for a typical sensing experiment. The resistance changes by about 6\% upon exposure to 150~ppm of methanol. (c) Plot of scaled \rv as a function of time for the same measurement as in (b).  In contrast to $R$, the noise changes by about 1500\% upon exposure to methanol. The blue and red blue dots represent the points in time when the time-series plotted in (a) were recorded.  \label{fig:timeseries}}}
\end {figure*}
\end{center}

In a typical sensing run, the chamber containing the device was evacuated and the values of both $R$ and \rv of the device were measured simultaneously to establish the base-line values. The device was then exposed to a known concentration of the chemical for a fixed period of time before the chamber was again evacuated. During this entire process both the resistance and the resistance fluctuations of the device were monitored simultaneously in real time.  A typical example of the enhancement in resistance fluctuations of the device upon exposure to chemical vapour is shown in figure~\ref{fig:timeseries}(a). The blue line shows the resistance  fluctuations for the pristine device while the red line is a plot of the resistance fluctuations after the device has been exposed to 150~ppm of methanol.
It can be seen from the time series that upon exposure to methanol the resistance fluctuations of the device is greatly enhanced. Figure \ref{fig:timeseries}(b) shows a plot of the \res and \rv measured simultaneously  during a typical sensing measurement. The values of both the parameters have been scaled by their respective values measured in the pristine device. During the first phase of the experiment (region I) the device was maintained in vacuum and the average $R$ and \rv were confirmed to be stable with time. Methanol was introduced into the measurement chamber at the instant of time marked by the the black dotted line. It was seen that  both  \res and \rv increased rapidly before saturating. The change in $R$ was about 6\%. On the other hand, the change in \rv was approximately 1500~$\%$, more than two orders of magnitude higher than the change in the resistance $R$.  At the end of phase II of the measurement the evacuation of the measurement chamber was started. The resistance takes a long time (of the order of few tens of minutes) to go back to its base-line value. This large recovery times of resistance based graphene sensors has been reported earlier by several authors~\cite{schedin2007detection,chen:053119,Wu2010296,6475143, :/content/aip/journal/apl/103/5/10.1063/1.4816762} and is a major hindrance in implementing resistance based graphene sensors. On the other hand the noise \rv reaches the base-line value within a few seconds. This very fast reset of \rv holds for a wide range of analyte concentrations and types.  Our measurements establish that a chemical sensor based on resistance fluctuations of the SLG-FET has at least two major advantages over a conventional resistance based detection scheme: 1) significantly higher sensitivity and (2) a much faster response.  

We have shown in a previous publication \cite{:/content/aip/journal/apl/106/18/10.1063/1.4919793} that that the most probable source of excess noise in these SLG-FET devices upon exposure to analytes is fluctuations in the number density of charge carriers in the SLG arising from adsorption-desorption of the chemicals at the device surface.  The values of the absorption-desorption energy ($E_a$) for several common analytes on the graphene surface is well known, both from theory and from experiments ~\cite{doi:10.1021/jp076538+,:/content/aip/journal/jap/112/6/10.1063/1.4752272,:/content/aip/journal/jcp/137/17/10.1063/1.4764356,doi:10.1021/ja403162r,:/content/aip/journal/jap/113/3/10.1063/1.4776239,doi:10.1021/jp8021024}. The presence of a definite energy scale $E_a$ associate with the adsorption-desorption of a specific gas on the graphene surface  gives rise to a characteristic frequency  scale $f_C$ in the measured $1/f$ noise spectrum. This characteristic frequency $f_C$ is directly related the absorption-desorption activation energy $E_a$ through the equation 
\begin{eqnarray}
f_C = f_0 exp \left(\frac{-E_a(T)}{k_BT}\right)
\label{eqn:fc}
\end{eqnarray}
where $f_0$ is the attempt frequency for the thermally activated process~
\cite{:/content/aip/journal/apl/106/18/10.1063/1.4919793,6475143}. The presence of a characteristic frequency $f_C$ in the measured $1/f$ spectrum is thus a spectroscopic signature specific to the analyte and allows a detection with specificity even in the presence of a mixture of gases~\cite{:/content/aip/journal/apl/106/18/10.1063/1.4919793}. 

\begin{center}
\begin {figure*}[!h]
    \includegraphics[width=0.75\textwidth]{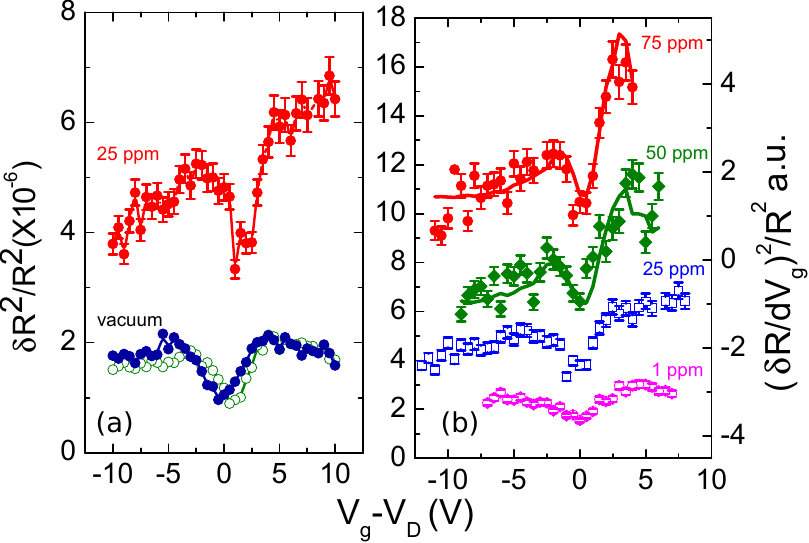}    
    \small { \caption{(a) Plots of \rv as a function of reduced gate voltage ($V_g-V_D$) of the pristine SLG-FET device (olive open circles), after introduction of 25~ppm of methanol to the measurement chamber (red filled circles) and after the methanol vapour has been pumped out (blue filled circles). (b) Plots of \rv as a function of reduced gate voltage ($V_g-V_D$) after the SLG-FET has been exposed to different amounts of methanol vapour - 1~ppm (magenta open circles), 25~ppm (blue open squares), 50~ppm (olive filled diamonds) and 75~ppm (red filled circles). (right axis) Plot of $1/R(dR/dV_g)^2$ for 50~ppm methanol (olive line) and 75~ppm methanol (red line).    
\label{fig:sensing_vg}}}
\end {figure*}
\end{center}

Intuitively, it appears that the sensing efficiency should be strongly correlated to the number density and to the nature of charge carriers present in the SLG. To test this we have carried out the sensing measurements at different values of back gate voltage ( $V_g$) - the results for a typical measurement are plotted in figure~\ref{fig:sensing_vg}(a). The green open squares are the relative variance \rv measured before the sensing experiment. The noise as a function of $V_g$ has the symmetric M-shape typical of high quality SLG devices. After each measurement the measurement chamber  was pumped out and the noise measured \sl{again} as a function of the gate voltage. It was seen that the noise in the device after the analyte was pumped out always came back to the base-line value of the pristine SLG-FET as shown by the blue filled circles in figure~\ref{fig:sensing_vg}(a).  

The red curve represents the noise \rv measured after the device has been exposed to 25~ppm of methanol. We note that there are two important features of the graph. First the noise is greatly enhanced at all values of $V_g$ in comparison to that in the pristine device. 
The second interesting feature is that the measured noise is no longer symmetric about the Dirac point - for a given amount of a certain analyte, the measured noise was seen to increase as the gate voltage was progressively made positive. In other words, the response of the graphene monolayer, when exposed to a fixed amount of analyte, was not symmetric in the electron-doped and in the hole-doped regimes - the response was much stronger in the electron doped regime ($(V_g - V_D)>0$) than in the hole-doped regime ($(V_g - V_D)<0$). The slope of the noise plots as a function of  the gate voltage at high values of $|V_g-V_D|$, where $V_D$ is the dirac point of graphene, was seen to scale almost linearly with the amount of analyte. This can be seen clearly in figure~\ref{fig:sensing_vg}(b) where we plot the noise as a function of $V_g$ in the presence of various concentrations of methanol. The evolution of the shape of the noise plots as  function of $V_g$ for different levels of exposure to analyte can be explained using the following analysis: if the dominant source of resistance noise in these devices upon exposure to the analyte is fluctuations in the number density $n$ of charge carriers in the conducting channel then the following relation holds: $\delta R = (\delta R/\delta n)\times \delta n$. Since the carrier concentration is controlled by the back gate voltage $V_g$, the contribution to resistance noise from number-density fluctuations would be \rv $\propto (dR/dV_g)^2/R^2$~\cite{kaverzin2012impurities}. The values of $(dR/dV_g)^2/R^2$ calculated from the measured $R-V_g$ curves are plotted on the right axis in figure~\ref{fig:sensing_vg}(b). There is very good qualitative match between the experimental data and the calculated plots. This shows that for SLG-FET devices exposed to chemicals number-density fluctuations is most probably the primary source of resistance fluctuation noise. 
\begin{center}
\begin {figure}[!tbh]
     \includegraphics[width=0.75\textwidth]{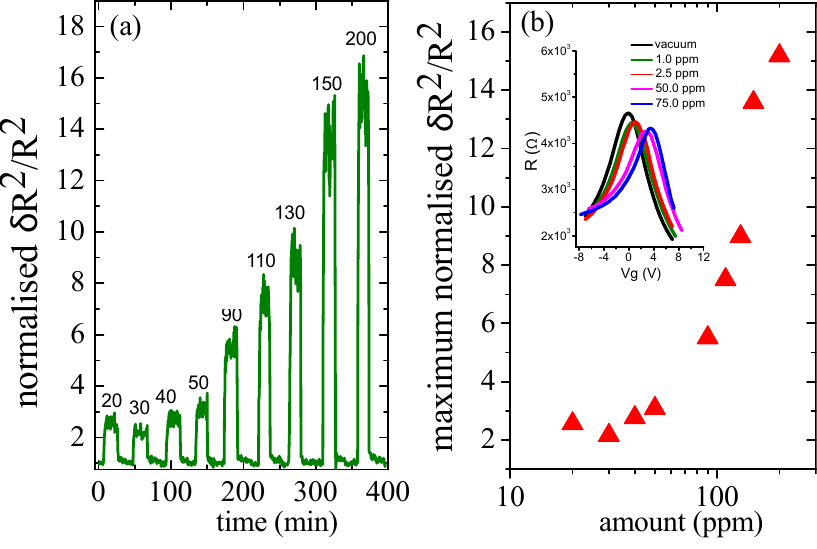}
    \small {\caption{ (a) Plot of the relative variance \rv of the SLG-FET device when exposed to different amounts of methanol vapour ranging from  20 ppm to 200 ppm. Very small response time and reset time was observed in all cases. (b) Plot of the average saturation value of \rv in presence of different concentrations of methanol vapour. The data  has been normalized by the baseline value of \rv before exposure to methanol. Inset: $R-V_g$ plots measured in the presence of different amounts of methanol vapour.\label{fig:amount_time}}}
\end {figure}
\end{center}

To test the scaling of the response of the SLG sensor with the concentration  of analytes, similar measurements as described above were performed by exposing the SLG-FET sensor to different amounts of chemicals.  The results obtained for methanol vapor ranging in concentration from 20~ppm to 300~ppm  are summarized in figure~\ref{fig:amount_time}.  Figure~\ref{fig:amount_time}(a)  shows the \rv with time for consecutive measurements carried out with different concentrations of methanol vapour. In the plot the values of \rv have been scaled by the baseline value of \rv measured in the pristine SLG device. Note that in each case the value of the relative variance \rv resets to initial state as the methanol vapour is pumped out. The average increase in \rv  after the device had been exposed to methanol normalized by the baseline value of \rv is plotted in figure \ref{fig:amount_time}(b). The relative variance of resistance fluctuations \rv was found to increase with the amount of analyte present in the measurement chamber - this scaling behavior was reproducible over several devices.

\begin{center}
\begin {figure}[!h]
  \includegraphics[width=0.5\textwidth]{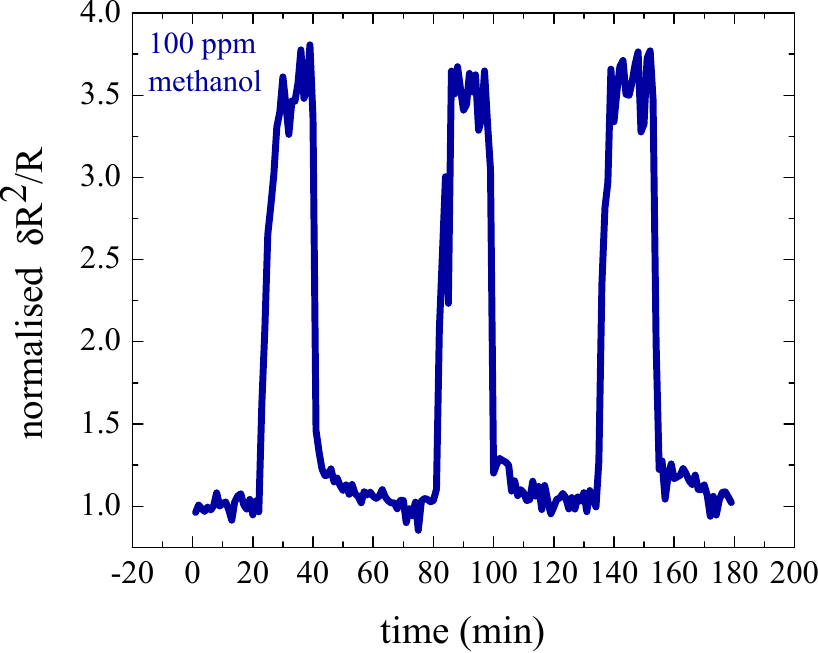}
    \small {\caption{ Plot showing the reproducibility in the change of  \rv of the SLG device for 3 consecutive runs with 100~ppm methanol. \label{fig:reprod}}}
\end {figure}
\end{center}

To test the reproducibility of our sensing scheme with noise, the SLG device was exposed to the same amount of methanol multiple times. After each exposure the device chamber was evacuated and it was ensured that the noise went down to the baseline value. A representative result is shown in figure~\ref{fig:reprod} where the normalized \rv is plotted as a function of time. From the plot it can be seen that every time the device was exposed to 100~ppm methanol vapour the relative variance of resistance fluctuations \rv scaled up to the same average value and upon pumping it sharply reset to the initial state. For the detection of 100~ppm methanol, the observed spread in the average value corresponds to an error in detection of 0.20~ppm attesting to the high fidelity of the sensor response.

\begin{center}
\begin {figure}[!h]
  \includegraphics[width=0.5\textwidth]{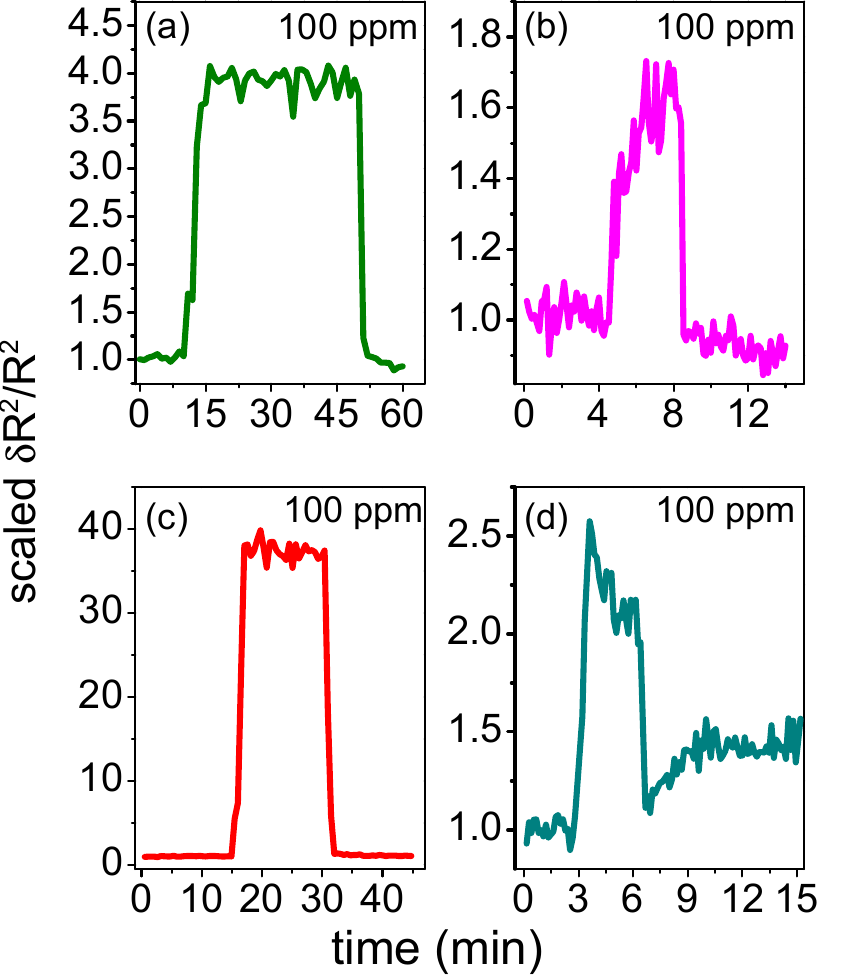}
    \small { \caption{Plots of \rv during sensing experiments with the SLG FET device exposed to 100~ppm of different chemicals (a) methanol, (b) chloroform, (c) nitrobenzene and (d) ammonia. \label{fig:diff_chem}}}
\end {figure}
\end{center}

Our sensing scheme based on noise measurements was tested using different types of molecules. The qualitative trend of fast response and high sensitivity was observed for all the chemicals tested. The quantitative  response of the \rv  to vapors of different chemicals varied, ranging from a $\sim~70~\%$ change for chloroform, to $\sim~300~\%$ change in the case of nitrobenzene. In figure \ref{fig:diff_chem} we plot  the response of the SLG sensor exposed to 100~ppm vapors of  methanol, ammonia, chloroform and nitrobenzene.  We were especially interested in the detection of nitrobenzene since nitro group chemicals are extensively used in explosives. In our measurements the \rv in graphene device was observed to be highly sensitive to presence of  nitrobenzene even at very low concentration making it a very promising sensor for detection of explosives. 

We currently do not have a clear understanding of the strong response of the SLG-FET sensor to nitrobenzene. It has been predicted that the  $NO_2$ functional group associated with nitrobenzene has a very strong affinity with graphene, which can result in strong scattering centers. A dynamic fluctuation of a strong scattering potential, resulting from the absorption-desprption of nitrobenzene,  might lead to the high noise levels measured.  We believe that further experimental and theoretical work is needed to address this issue.

\section{CONCLUSION}

To conclude, we present experiments testing the efficacy of graphene monolayer FET based devices as chemical sensors. We find that a detection scheme based on the measurements of resistance fluctuations is far superior to the traditional method of measuring the average resistance change in terms of sensitivity, specificity and the response time of the detector. We show that the for monolayer graphene devices exposed to the ambient the most likely source of enhanced resistance fluctuations is fluctuations in the carrier number density. To the best of our knowledge graphene based chemical sensors with these characteristics have not been reported before. 

\section{ACKNOWLEDGEMENT}

The authors thank  NPMASS, Govt. of India for support.  The authors thank device fabrication and characterization facilities at  National Nanofabrication Centre and Micro and Nano Characterization Facility in Centre for Nano Science and Engineering at IISc, Bangalore. KRA thanks  CSIR, MHRDG, Govt. of India for support.

\bibliography{sensor}

\end{document}